\documentclass[amsfonts,12pt]{article}

\usepackage[dvips]{graphicx}

\makeatletter
\def\@cite#1#2{{[{#1}]\if@tempswa\typeout
{IJCGA warning: optional citation argument
ignored: `#2'} \fi}}

\newcount\@tempcntc
\def\@citex[#1]#2{\if@filesw\immediate\write\@auxout{\string\citation{#2}}\fi
  \@tempcnta\z@\@tempcntb\m@ne\def\@citea{}\@cite{\@for\@citeb:=#2\do
    {\@ifundefined
       {b@\@citeb}{\@citeo\@tempcntb\m@ne\@citea\def\@citea{,}{\bf ?}\@warning
       {Citation `\@citeb' on page \thepage \space undefined}}%
    {\setbox\z@\hbox{\global\@tempcntc0\csname b@\@citeb\endcsname\relax}%
     \ifnum\@tempcntc=\z@ \@citeo\@tempcntb\m@ne
       \@citea\def\@citea{,}\hbox{\csname b@\@citeb\endcsname}%
     \else
      \advance\@tempcntb\@ne
      \ifnum\@tempcntb=\@tempcntc
      \else\advance\@tempcntb\m@ne\@citeo
      \@tempcnta\@tempcntc\@tempcntb\@tempcntc\fi\fi}}\@citeo}{#1}}
\def\@citeo{\ifnum\@tempcnta>\@tempcntb\else\@citea\def\@citea{,}%
  \ifnum\@tempcnta=\@tempcntb\the\@tempcnta\else
   {\advance\@tempcnta\@ne\ifnum\@tempcnta=\@tempcntb \else \def\@citea{--}\fi
    \advance\@tempcnta\m@ne\the\@tempcnta\@citea\the\@tempcntb}\fi\fi}
\makeatother
\newenvironment{Eqnarray}%
     {\arraycolsep 0.14em\begin{eqnarray}}{\end{eqnarray}}

\def\be{\begin{equation}}
\def\ee{\end{equation}}
\def\bear{\be\begin{array}}
\def\eear{\end{array}\ee}
\def\bea{\begin{Eqnarray}}
\def\eea{\end{Eqnarray}}

\def\lsim{\mathrel{\raise.3ex\hbox{$<$\kern-.75em\lower1ex\hbox{$\sim$}}}}
\def\gsim{\mathrel{\raise.3ex\hbox{$>$\kern-.75em\lower1ex\hbox{$\sim$}}}}
\def\ifmath#1{\relax\ifmmode #1\else $#1$\fi}
\def\ls#1{\ifmath{_{\lower1.5pt\hbox{$\scriptstyle #1$}}}}

\def\beq{\begin{equation}}
\def\eeq{\end{equation}}
\def\beqa{\begin{Eqnarray}}
\def\eeqa{\end{Eqnarray}}

\def\boxit#1{\leavevmode\thinspace\hbox{\vrule\vtop{\vbox{\hrule%
        \vskip3pt\kern1pt\hbox{\vphantom{\bf/}\thinspace\thinspace%
        {\bf#1}\thinspace\thinspace}}\kern1pt\vskip3pt\hrule}\vrule}%
        \thinspace}
\def\Boxit#1{\noindent\vbox{\hrule\hbox{\vrule\kern3pt\vbox{
        \advance\hsize-7pt\vskip-\parskip\kern3pt\bf#1
        \hbox{\vrule height0pt depth\dp\strutbox width0pt}
        \kern3pt}\kern3pt\vrule}\hrule}}

\def\stau{{\widetilde\tau}}
\def\smu{{\widetilde\mu}}
\def\se{{\widetilde e}}

\def\baselinestretch{1}
\begin{document}

\catcode`@=11
\newtoks\@stequation
\def\subequations{\refstepcounter{equation}%
\edef\@savedequation{\the\c@equation}%
  \@stequation=\expandafter{\theequation}
  \edef\@savedtheequation{\the\@stequation}
  \edef\oldtheequation{\theequation}%
  \setcounter{equation}{0}%
  \def\theequation{\oldtheequation\alph{equation}}}
\def\endsubequations{\setcounter{equation}{\@savedequation}%
  \@stequation=\expandafter{\@savedtheequation}%
  \edef\theequation{\the\@stequation}\global\@ignoretrue

\noindent}
\catcode`@=12
\setcounter{footnote}{1} \setcounter{page}{1}

\noindent

\title{{\bf Probing lepton flavour violation\\ in slepton NLSP
scenarios}}
\vskip2in
\author{    
{\bf Koichi Hamaguchi$^{1}$\footnote{\baselineskip=16pt E-mail: {\tt
koichi.hamaguchi@desy.de}. }} and  
{\bf Alejandro Ibarra$^{2,3}$\footnote{\baselineskip=16pt  E-mail: {\tt
alejandro.ibarra@cern.ch}}} \\ 
\hspace{3cm}\\
 $^{1}$~{\small Deutsches Elektronen-Synchrotron, DESY} \\
{\small 22603 Hamburg, Germany}
\hspace{0.3cm}\\
 $^{2}$~{\small Department of Physics, Theory Division, CERN}\\
{\small CH-1211 Geneva 23, Switzerland}
\hspace{0.3cm}\\
 $^{3}$~{\small Instituto de F\'{\i}sica Te\'orica, CSIC/UAM, C-XVI} \\
{\small Universidad Aut\'onoma de Madrid,} \\
{\small Cantoblanco, 28049 Madrid, Spain.}
}
\maketitle

\def\baselinestretch{1.15}
\begin{abstract}
\noindent
In supersymmetric models where the gravitino is the lightest
superparticle, the next-to-lightest superparticle (NLSP) is
long-lived, and hence could be collected and studied in detail. We
study the prospects of direct detection of lepton flavour violation in
charged slepton NLSP decays. 
Mixing angles in the slepton sector as small as 
$\sim 3\times 10^{-2}~(9\times 10^{-3})$ could be probed at the 90\% 
confidence level
if $3\times 10^3$ ($3\times 10^4$) sleptons could be collected.
\end{abstract}

\thispagestyle{empty}

\vskip-17cm
\rightline{CERN-PH-TH/2004-247}
\rightline{DESY-04-231}
\rightline{IFT-UAM/CSIC-04-61}

\newpage
\baselineskip=20pt

\section{Introduction}

In a supersymmetric (SUSY) theory with R-parity
conserved the lightest superparticle (LSP) is stable.
This property puts forward the LSP as a natural candidate for
the dark matter of the Universe, and several candidates
have been proposed in the literature: neutralinos, 
axinos, gravitinos...
In this paper we concentrate on the darkest among all the dark matter
candidates, the gravitino, 
which in the early universe can be produced 
by thermal scatterings~\cite{gravitino-thermal}
and in the decays of superparticles~\cite{gravitino-NLSP}.

If the gravitino is the LSP, the next-to-lightest superparticles
(NLSPs) can only decay to gravitinos and other Standard Model particles
via gravitational interactions.  As a consequence, NLSPs are very long
lived and could be collected and studied with detail in future
experiments.  Recently, prospects of collecting charged NLSPs and
detecting their decay products in future colliders have been 
discussed~\cite{Hamaguchi:2004df,Feng:2004yi}.  At the LHC, cascade
decays of squarks and gluinos can produce of the order of $10^6$ NLSPs
per year if the sparticle masses are close to the current experimental
limits \cite{Beenakker:1996ch}.
Among them ${\mathcal O}(10^3$--$10^4)$ NLSPs could be collected by
placing 1--10 kton massive material around the LHC detectors.  On the
other hand, at the Linear Collider~\cite{Aguilar-Saavedra:2001rg} up
to ${\mathcal O}(10^3$--$10^5)$ NLSPs may be collected and studied.
If the stopper material simultaneously serves as an active real-time
detector~\cite{Hamaguchi:2004df}, the decay products and their
distributions could be studied in detail there.

If a sample of NLSPs could be accumulated, their decays could be
observed and studied without important backgrounds, allowing a precise
determination of the supersymmetric parameters.  In
Ref.~\cite{Buchmuller:2004rq}, NLSPs decays were used as a method to prove
the existence of supergravity in nature. Measurement of the NLSP
lifetime, together with kinematically reconstructed gravitino mass,
could lead to a microscopic determination of the Planck scale. Moreover,
study of a rare three-body decay could reveal the peculiar couplings of
gravitino, and may even determine the gravitino spin.

Future colliders will open new opportunities to detect
lepton flavour violation (LFV), complementary to the
ongoing and projected experiments on rare decays. 
There is mounting literature on the prospects to detect 
LFV in the LHC or in the future Linear Collider \cite{LFV-LC}, under 
the implicit assumption that the neutralino is the LSP.
In this paper we assume that the gravitino is the LSP and
the NLSP is a charged slepton, and study
the prospects of direct detection of lepton flavour violation  in
their decays.  The basic idea is very simple. Suppose that many NLSP
sleptons could be collected, say staus. If there is no LFV, 
all of them would decay into taus and gravitinos, 
$\stau\to\tau\; \psi_{3/2}$. If on the contrary LFV exists in
nature, some of the collected staus would decay as $\stau\to
e\; \psi_{3/2}$ or $\stau\to\mu\; \psi_{3/2}$, which would 
constitute a direct signal of LFV.

Different scenarios will arise depending on 
the spectrum of charged sleptons
and on their different decay modes. This is discussed in detail
in Section 2. The signals for lepton flavour violation
in NLSP decay are systematically studied for the different
scenarios in Section 3. Our conclusions are presented in 
Section 4. Finally, we include two appendices with a review
of the electron spectrum from the decay of taus in flight, 
and some comments about the implementation of this idea for
the case with R-parity violation.

\section{Charged slepton mass spectrum and decays}

The charged slepton masses receive contributions from the supersymmetric
Lagrangian and from the soft breaking terms. The supersymmetric
contributions include those from the superpotential,
that are related to the charged lepton masses, and the D-terms.
In the minimal scenario, the D-term contribution is proportional
to the $Z$ mass and $\tan\beta$, although in 
more general scenarios there could be additional 
contributions coming from the breaking of extra
gauge groups at intermediate energies, that could be 
flavour conserving or flavour violating \cite{Drees:1986vd}. In addition to 
these, there is a contribution to the charged slepton masses from 
the SUSY soft breaking terms:
\bea
-{\cal L}_{\rm soft}=
(m^2_{\widetilde l_L})_{ij} {{\widetilde l_{L_i}}^{\dagger}} 
{\widetilde l_{L_j}} +
(m^2_{\widetilde l_R})_{ij} {{\widetilde l_{R_i}}^{\dagger}} 
{\widetilde l_{R_j}} +
(Y^l_{ij} {A_l}_{ij} {\widetilde l_{R_i}}{\widetilde l_{L_j}} H_d  +h.c.),
\eea
where $i,j=1,2,3$ are generational indices. ${\widetilde l_{L,R}}$ denote
the left and right-handed charged sleptons, 
$H_d$ is the down-type Higgs doublet,
$Y^l_{ij}$ is the charged-lepton Yukawa coupling,
and $(m^2_{\widetilde l_{L,R}})_{ij}$ and ${A_l}_{ij}$ are the soft
scalar mass matrix squared and the soft trilinear matrix, respectively.
This contribution is in general flavour violating and mixes left-handed
and right-handed sleptons through the trilinear terms.

The resulting $6\times 6$ mass matrix reads:
\bea
{\cal M}^2=
\pmatrix{m^2_{\widetilde l_L} + m^2_{l} 
-(\frac{1}{2}-\sin^2 \theta_W)m^2_Z \cos 2\beta &
m_l(A_l-\mu \tan \beta) \cr
m_l(A_l-\mu \tan \beta) &
m^2_{\widetilde l_R} + m^2_{l} -\sin^2 \theta_W m^2_Z \cos 2\beta},
\eea
which is diagonalized by certain mass eigenstates, 
${\widetilde l_{\alpha}}$. We will label the mass eigenstates
so that ${\widetilde l_{1}}$ is the lightest and ${\widetilde l_{6}}$
the heaviest. The mass eigenstates are
related to the flavour eigenstates by the unitary transformation
\bea
{\widetilde l_{\alpha}}=U_{\alpha,i} {\widetilde l_{L_i}}
+U_{\alpha,i+3}{\widetilde l_{R_i}},
\eea
where  $\alpha=1,... 6$ runs over
the mass eigenstates, and $i=1,2,3$ runs over the flavour eigenstates 
(e, $\mu$, $\tau$). For instance, the matrix 
element $U_{\alpha, 3}$ measures the ${\widetilde \tau_L}$ 
content of the mass  eigenstate ${\widetilde l_{\alpha}}$, 
whereas $U_{\alpha, 6}$ indicates the ${\widetilde \tau_R}$ content. 

If flavour violation is not very large in nature, 
as present experiments suggest, then the mass eigenstates are mainly 
a combination of left and right-handed sfermions of
the same generation. We expect that the lightest charged slepton
is mainly a combination of staus, as a result of the effect of 
the trilinear terms and of the negative contribution to the slepton masses 
from the tau radiative
corrections. We will assume this in what follows,
although our results can be generalized to the other possibilities. 

{}From the mass matrix, it is apparent that when the left and right-handed
soft scalar masses are approximately equal at high energies, 
$m^2_{\widetilde l_L} \simeq m^2_{\widetilde l_R}$,
the right-handed sfermions are going to be 
lighter than their left-handed counterparts, 
due to the D-term contribution to the scalar masses. There is however
a more important effect that splits left and right handed sleptons,
namely radiative corrections. Left-handed sfermion masses 
receive a positive contribution at low energies from the wino 
radiative corrections, whereas the right-handed
ones do not. Therefore, in a scenario where at high energies the left-handed
and right-handed soft masses are very similar,
the lightest charged slepton will approximately 
correspond to the right-handed stau, 
i.e., ${\widetilde l_1}\approx {\widetilde \tau_R}$. Nevertheless, when
$\tan \beta$ is large, ${\widetilde l_1}$ can
have a non-negligible component of left-handed stau,
because of the enhancement of left-right mixing at large $\tan \beta$,
that otherwise is suppressed by the small tau mass. 
In smuon and selectron sectors left-right mixings are very small, and
the (next-to-) next-to-lightest charged slepton will
almost correspond to the right-handed smuon (selectron), 
i.e., ${\widetilde l_2},\,{\widetilde l_3}
\approx {\widetilde \mu_R},\,{\widetilde e_R}$.

In the presence of lepton flavour violations the mass eigenstates are
in general linear combinations of all the flavour eigenstates. For
simplicity, in the following the mass eigenstates will be sometimes
referred by the closest flavour eigenstates. For instance,
$m_{\stau}$ will represent the mass of ${\widetilde l_1}$, which
is mainly the stau ${\stau}$.

After discussing qualitatively the features of the charged slepton
spectrum, let us study their decay modes in the scenario
where the gravitino is the LSP. 
Charged sleptons can decay
via gravitational interactions to a gravitino and a 
charged lepton. If the lightest charged slepton
is the NLSP, this is the only possible decay channel.
On the other hand, heavier charged sleptons can also decay
to NLSPs and neutrinos via chargino exchange
${\widetilde l^-_2} \rightarrow {\widetilde l^-_1} 
\;{\bar \nu_{i}}\; \nu_{j}$, or to NLSPs and charged 
leptons via neutralino exchange
${\widetilde l^-_2} \rightarrow {\widetilde l^-_1} 
\;{\bar l^{\pm}_{i}}\; l^{\mp}_{j}$  (and analogously for
the positively charged sleptons). 
These processes and their
decay rates were studied in detail in~\cite{Ambrosanio:1997bq}.

The decay of a charged slepton mass eigenstate
into a gravitino and a charged lepton is described
by the following Lagrangian~\cite{WessBagger}:
\begin{eqnarray}
  {\cal L}_{3/2}=-\frac{1}{\sqrt{2} M_P} \left[ 
    U^*_{\alpha,i} 
    (D_{\nu} {\widetilde l_\alpha}) {\bar l_i} \; P_R
    \gamma_{\mu} \gamma^{\nu}  \psi^{\mu}+
    U^*_{\alpha,i+3} 
    (D_{\nu} {\widetilde l_\alpha}) {\bar l_i} \;P_L
      \gamma_{\mu} \gamma^{\nu} \psi^{\mu}+ h.c. 
      \right],
\end{eqnarray}
where $D_{\nu} {\widetilde l_{\alpha}}= (\partial_\nu + i e A_\nu)
{\widetilde l_{\alpha}}$ is the covariant derivative and
$M_P= (8 \pi G_N)^{-1/2}$ is the reduced Planck mass.
From this Lagrangian, it is straightforward to compute the total
decay rate of the slepton ${\widetilde l_{\alpha}}$ into 
a lepton with flavour $i$ and a gravitino. The result is
\bea
&&\Gamma({\widetilde l_{\alpha}}\rightarrow l_i \; \psi_{3/2})
\nonumber\\
&=& 
\frac{m^5_{\widetilde l_{\alpha}}}{48 \pi m^2_{3/2} M^2_P}
\left( |U_{\alpha, i}|^2+ |U_{\alpha, i+3}|^2 \right)
\left(1-\frac{m^2_{3/2}+m_{l_i}^2}{m^2_{\widetilde l_{\alpha}}}\right)^4
\nonumber\\
&&\times
\left(1
-\frac{4 m^2_{3/2} m_{l_i}^2}
{(m^2_{\widetilde l_{\alpha}}-m_{3/2}^2-m_{l_i}^2)^2}\right)^{3/2} 
\left(
1
-\frac{4m_{3/2}m_{l_i}}{m^2_{\widetilde l_{\alpha}}-m_{3/2}^2-m_{l_i}^2}
\cdot
\frac{\mathrm{Re}U_{\alpha, i}U_{\alpha, i+3}^*}
{|U_{\alpha, i}|^2+ |U_{\alpha, i+3}|^2}\right)
\nonumber \\
&\simeq& 
\frac{m^5_{\widetilde l_{\alpha}}}{48 \pi m^2_{3/2} M^2_P}
\left( |U_{\alpha, i}|^2+ |U_{\alpha, i+3}|^2 \right)
\left(1-\frac{m^2_{3/2}}{m^2_{\widetilde l_{\alpha}}}\right)^4,
\label{decay}
\eea
where in the last step we have neglected the mass of the lepton.
The decay rates are typically of the order of $10^{-22}$GeV
to $10^{-33}$GeV for gravitino masses of 
$\sim 1$MeV to $\sim 100$GeV, and NLSP masses of $\sim150$GeV. 
In the case of the lightest charged slepton, that can only
decay to a gravitino, these decay rates translate into 
lifetimes of the order of mili-seconds to years, respectively.

The key idea of the present analysis comes from the fact
that these decays could have a
non-trivial flavour structure, that is reflected in the 
$6 \times 6$ unitary matrix $U$. Therefore, one could directly
measure the slepton mixings by studying the branching
ratios:
\bea
|U_{\alpha, i}|^2+ |U_{\alpha,i+3}|^2 
\propto
BR({\widetilde l_{\alpha}}\rightarrow l_i \; \psi_{3/2}).
\eea
Especially,
\bea
|U_{1, i}|^2+ |U_{1,i+3}|^2 
\simeq
BR({\widetilde l_1}\rightarrow l_i \; \psi_{3/2}).
\eea

For this to be feasible, it would be desirable that all the heavier
sleptons decay fast enough into the NLSP.\footnote{In the following
we will assume that neutralinos are heavier than the next-to-lightest
charged sleptons ${\widetilde l_2},\,{\widetilde l_3} \approx
{\widetilde \mu_R},\,{\widetilde e_R}$.  If on the contrary the
lightest neutralino ${\widetilde N_1}$ is lighter than the selectron
(smuon), i.e., $m_{\widetilde \tau}<m_{\widetilde N_1} < m_{\widetilde
e_R} (m_{\widetilde \mu_R})$, the ${\widetilde e_R}$ (${\widetilde
\mu_R}$) mainly decays into ${\widetilde N_1}$, not into ${\stau}$.
However, as far as $m_{\stau}+m_{\tau} < m_{\widetilde N_1} <
m_{\widetilde e_R} (m_{\widetilde \mu_R})$, the neutralino
${\widetilde N_1}$ quickly decays into $\stau$ and hence reduces to
the cases discussed in the text. If the neutralino mass happens to lie
in the narrow range $m_{\stau} < m_{\widetilde N_1} <
m_{\stau}+m_{\tau}$, ${\widetilde N_1}$ can have a long lifetime and
may escape from the stopper material.} Otherwise, the stopping
material would be polluted with next-to-NLSPs (NNLSPs), whose decays
could represent sources of background for the lepton flavour violating
decays that we want to study. We will show that if lepton flavour violation
is large enough to be observed experimentally, it is guaranteed
that this situation will not happen, allowing a clean observation
of the LFV effects.

For the sake of the presentation, let us first discuss 
the decay modes in the case without flavour violation.
The dominant decay channel is normally the one mediated
by neutralinos, provided that the decay is kinematically 
accessible. This requires a mass difference 
between the NLSP and the NNLSP larger than the tau mass,
that despite being very small is not always guaranteed.
For instance, radiative corrections induced by the tau 
generate a mass difference which is 
$m_{\widetilde l_2}-m_{\widetilde l_1} \sim 10^{-2} m_{\tau} \tan^2\beta$, 
that is large enough
when $\tan \beta$ is large. Moreover, left-right mixing 
also produces a mass splitting between the NLSP and the NNLSP,
that reads:
\bea
m_{\widetilde l_2}-m_{\widetilde l_1} \simeq \frac{m^2_{\tau}(A_{\tau}-\mu  \tan\beta)^2}
{2m_{\widetilde l_1}\left(m^2_{\widetilde l_L}-m^2_{\widetilde l_R}+
(2\sin^2\theta_W-\frac{1}{2})m_Z^2 \cos2\beta \right)}.
\eea
Hence, $m_{\widetilde l_2}-m_{\widetilde l_1}>m_{\tau}$ is obtained when
$|A_{\tau}-\mu  \tan\beta| \gsim 
\sqrt{2m_{\widetilde l_1}
(m^2_{\widetilde l_L}-m^2_{\widetilde l_R})/m_{\tau}}$, 
that is fulfilled for some values of
$A_{\tau}$, or for large values of $\tan\beta$. 

\begin{figure}[t]
  \centering
  \includegraphics[width=12cm,clip=true]{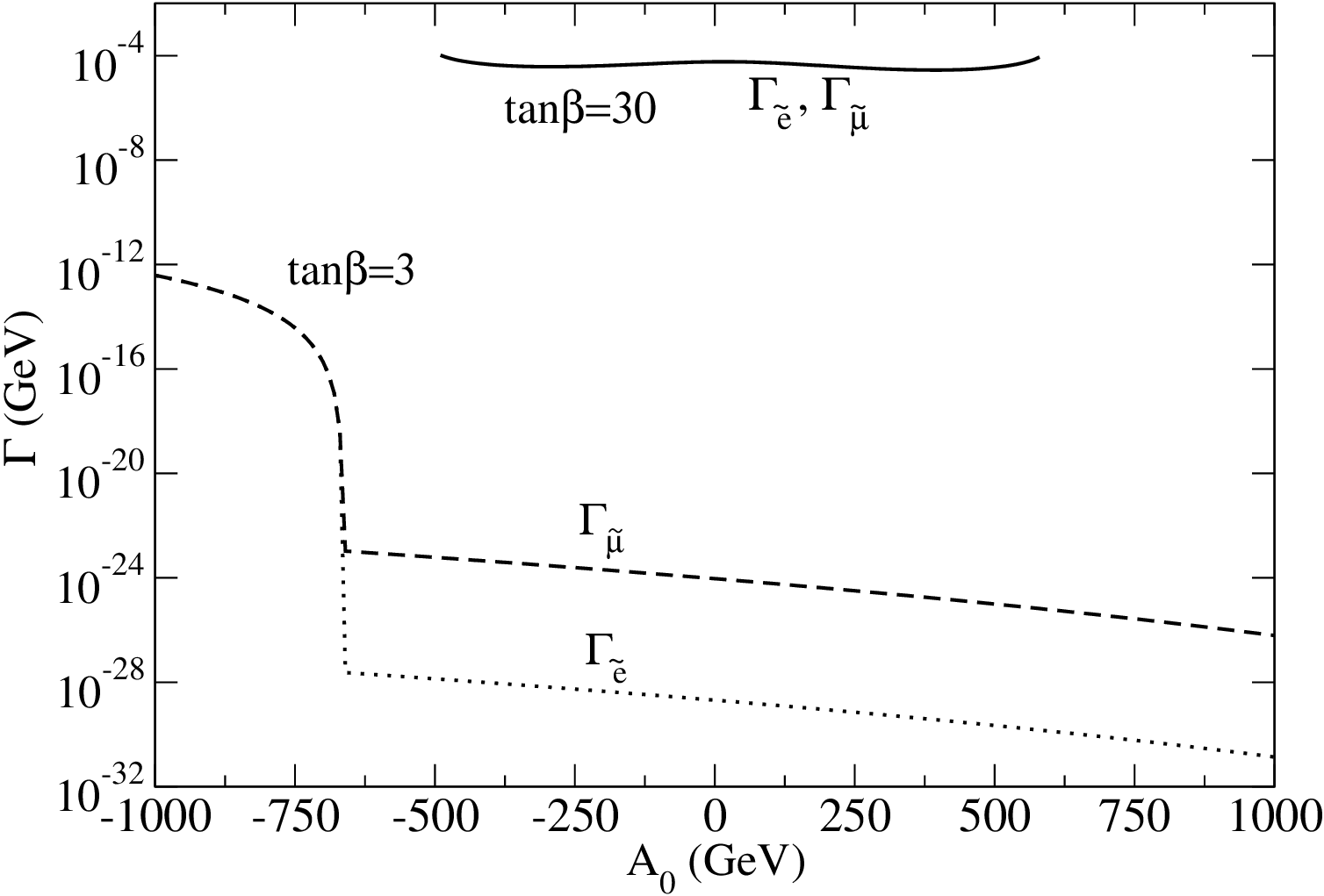}
  \caption{Decay rates for the selectron and the smuon for specific
  choices of SUSY parameters in the constrained MSSM scenario,
  namely, $m_0=0$, $M_{1/2}=400 {\rm GeV}$, $\mu>0$, for the
  case without lepton flavour violation. These parameters
  lead to charged slepton masses of 
  $m_{\widetilde l_1}\approx 150~{\rm GeV}$ for
  $\tan\beta=3$ and $m_{\widetilde l_1}\approx 50-100~{\rm GeV}$ for
  $\tan\beta=30$. For large 
  $\tan\beta$ and large $|A_0|$ the spectrum presents tachyonic
  sleptons.}
  \label{fig-gamma}
\end{figure}

Therefore, when $\tan \beta$ is large, generically larger than $\sim
10$, the mass difference between the NLSP and the NNLSP is large
enough to allow the decay via neutralino exchange. On the other hand,
when $\tan \beta$ is small, there are regions of the parameter space
where this mass difference is not large enough and the NNLSP can only
decay into the NLSP via chargino exchange.\footnote{We will neglect
the 4- and 5-body decay modes via virtual tau exchange, $\se\to
\stau\; e\; \bar{\nu}_\tau \;X$ where $X$ is $e\;\bar{\nu}_e$,
$\mu\;\bar{\nu}_\mu$, $\pi^-$ and so on. The rates of those modes may
become comparable to $\Gamma(\se\to\stau\;\bar{\nu}_\tau\; \nu_e)$. In
such cases, $\Gamma(\se\to\stau\;\bar{\nu}_\tau\;\nu_e)$ should be
replaced with $\Gamma(\se\to\stau\;\bar{\nu}_\tau\; \nu_e)+\Gamma(\se\to
\stau\; e\;\bar{\nu}_\tau \;X)$ in the following discussion, but our
conclusions are not affected.}
The latter decay channel is
suppressed by small Yukawa couplings, resulting in very small decay
rates. In particular, the decay rate of the right-handed selectron via
chargino interaction can be so small, that the decay into gravitinos
could become the dominant channel.  If this happens, the sample would
be polluted with right-handed selectrons, and the electrons from their
flavour conserving decay 
${\se}\to e\;\psi_{3/2}$ 
could mask the electrons from the flavour
violating decay of the staus
${\stau}\to e\;\psi_{3/2}$.  
The decay rates for the
NNLSP and the NNNLSP are illustrated in Fig. \ref{fig-gamma}, 
for a specific choice of SUSY parameters compatible with 
cosmology~\cite{cosmology}.
In the plot, we have omitted the contribution from the 
decays into gravitino. 
We assume the boundary conditions for the
constrained MSSM, and we run the renormalization group equations to
low energies to obtain the mass spectrum. As usual, the $\mu$ parameter is
determined by imposing a correct electroweak symmetry
breaking.
For $\tan\beta=3$, the decay of smuon (selectron)
into stau, tau and a charged lepton 
via neutralino exchange is kinematically forbidden
for $A_0\gsim -650$GeV, and the decay rates become suppressed. For
larger $\tan\beta$, the decay into tau is open and the rates become
large.

It is interesting to note that if flavour violation exists
in nature this picture changes radically. The
reason for this is double. First, flavour violation in the right-handed
sector can induce a non-degeneracy among the mass eigenstates, 
that can be large enough to open kinematically the
(fast) decay channel mediated by neutralinos. To be precise,
this occurs when 
$(m^2_{\widetilde l_R})_{13}/m_{\widetilde l_1} \gsim m_{\tau}$.
Analogously, some amount of flavour violation in the
right-handed smuon-stau sector is enough to open
kinematically the decay $\smu \rightarrow \stau \; \mu \;\tau$.
Notice that $(m^2_{\widetilde l_R})_{23}/m_{\widetilde l_1} \gsim m_{\tau}$
is not in conflict with the present
bound $BR(\tau \rightarrow \mu\; \gamma) \lsim 6 \times 10^{-7}$,
that requires $(m^2_{\widetilde l_R})_{23}/m_{\widetilde l_1} \lsim 100$GeV
for $\tan \beta \simeq 3$ and $m_{\widetilde l_1}\simeq 150$GeV 
\cite{Masina:2002mv}.

\begin{figure}[t]
  \centering
  \includegraphics[width=12cm,clip=true]{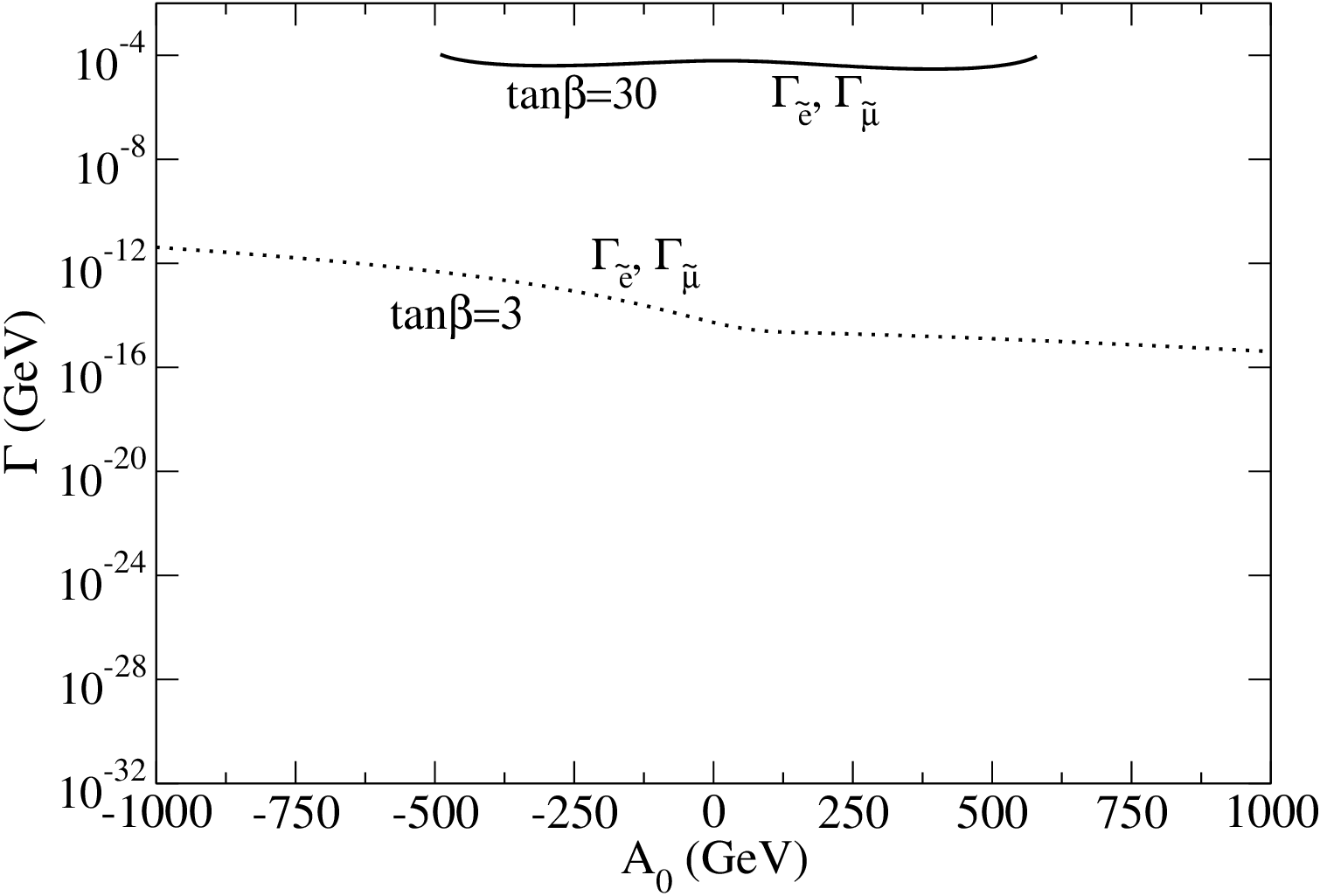}
  \caption{The same as fig.\ref{fig-gamma} with a small
    flavour violation in the right-handed slepton mass matrix:
    $(m^2_{\widetilde l_R})_{23}/m_{\widetilde l_1}^2=
    (m^2_{\widetilde l_R})_{13}/m_{\widetilde l_1}^2=0.01$.
  }
  \label{fig-gammaLFV}
\end{figure}

There is nevertheless a more important effect of flavour violation.
If flavour violation exists in nature, even in very small amounts,
the flavour violating decay channel mediated by neutralinos
$\se \rightarrow \stau \;e \;e$ can be very efficient. Despite the
decay rate is suppressed by the small mixing angle, it
could be orders of magnitude larger than the decay rates
of the processes mediated by charginos or into gravitinos.
Notice also that this decay channel is 
usually kinematically open, since
radiative corrections already induce a mass splitting between the
selectron and the stau larger than twice the electron mass.
In Fig. \ref{fig-gammaLFV}, we show the 
effect of flavour violation on the decay
rates of the smuon and the selectron.
It is remarkable the double role that lepton flavour
violation plays in our analysis: it is not only the 
object of our investigation, but also 
a crucial ingredient for the preparation of a sample with just NLSPs,
that would allow a clean detection of the flavour violation itself.

\section{Possible scenarios}
\label{section3}

Whether the sample contains just NLSPs or on the contrary
also contains (N)NNLSPs depends on the relative size
of the different decay rates (gravitational and mediated
by charginos and neutralinos).
Let us consider the case in which
only two species are relevant, namely staus and selectrons. 
We assume that smuons are not present in the sample, since
in general  these will decay fast enough into staus. Notice
from Fig.~\ref{fig-gamma} that even the decay by chargino exchange,
the most inefficient one apart from the gravitational channel,
is generically orders of magnitude larger than the 
decay into gravitinos.\footnote{We discard 
the extreme case where $\Gamma(\smu\to
{\stau}\;\nu_{\mu}\;\bar{\nu}_{\tau})\sim 10^{-24}\mathrm{GeV}
\lsim \Gamma(\smu\to \mu \; \psi_{3/2})$, which can happen
if $\tan\beta$ is small and 
$m_{3/2}\lsim 3~\mathrm{MeV}(m_{\smu}/100\mathrm{GeV})^{5/2}$.}
However, as we mentioned
in the previous section, this is not always the 
case for the selectrons, whose decay rate via chargino exchange is 
further suppressed by the small electron Yukawa
coupling. 

In a sample that contains just stopped selectrons and staus, the following 
decay processes can occur. There are the lepton flavour
conserving decays into gravitinos,
\bea
\Gamma({\stau}\rightarrow \tau \;\psi_{3/2}) 
\equiv \Gamma^{\stau}_{FC}\;, \nonumber \\
\Gamma({\se}\rightarrow e \;\psi_{3/2}) 
\equiv \Gamma^{\se}_{FC}\;;
\eea
the decays of the selectron into staus via neutralino or
chargino exchange, both the flavour conserving channel
and possibly also a flavour violating channel,
\bea
\Gamma({\se}\rightarrow {\stau}\; {\bar \nu_{\tau}}\; \nu_e)
+\Gamma({\se}\rightarrow {\stau}\; \tau\; e)
&\equiv& \Gamma_{FC,{\se}{\stau}}\;,  \nonumber\\
\Gamma({\se}\rightarrow {\stau}\; e\; e)
&\equiv& \Gamma_{FV,{\se}{\stau}}\;;
\eea
and possibly lepton flavour violating decays into gravitinos
\bea
\Gamma({\stau}\rightarrow e\; \psi_{3/2} )
\equiv \Gamma^{\stau}_{FV,e}\;, \nonumber\\
\Gamma({\stau}\rightarrow \mu\; \psi_{3/2} )
\equiv \Gamma^{\stau}_{FV,\mu}\;, \nonumber\\
\Gamma({\se}\rightarrow \tau\; \psi_{3/2})
\equiv \Gamma^{\se}_{FV,\tau}\;, \nonumber\\
\Gamma({\se}\rightarrow \mu\; \psi_{3/2})
\equiv \Gamma^{\se}_{FV,\mu} \;.
\eea
Here, we have omitted the selectron-smuon conversion 
$\se\to \smu\;\bar{\nu}_\mu\;\nu_e$ 
(or $\to\smu\; \mu \;e$) assuming that it is much suppressed 
compared to the selectron-stau conversion.\footnote{Even if 
$\se\to \smu\;\bar{\nu}_\mu\;\nu_e$ 
(or $\to\smu \;\mu  \;e$) occurs, 
it is followed by a prompt decay of smuon,
$\smu\to \stau\;\bar{\nu}_\tau \;\nu_\mu$ 
(or $\to\stau\;\tau\;\mu$), and effectively the whole process 
plays the role of ${\se}\to {\stau}$ conversion.}

The total decay rates for the stau and the selectron read:
\bea
\Gamma^{\stau}_{\rm total}&=&\Gamma^{\stau}_{FC}+\Gamma^{\stau}_{FV}\;,
   \nonumber \\
  \Gamma^{\se}_{\rm total}
  &=& \Gamma^{\se}_{FC}+\Gamma^{\se}_{FV}
  +\Gamma_{{\se}{\stau}}\;,
\eea
where we have defined $\Gamma_{{\se}{\stau}}\equiv
\Gamma_{FC,{\se}{\stau}}+\Gamma_{FV,{\se}{\stau}}$ as the total
decay rate of conversion of selectron into stau.

The abundances of the different species are determined by the
following set of differential equations:
\bea
\frac{dN_{\stau}}{dt}&=&-\Gamma^{\stau}_{\rm total}
N_{\stau}+\Gamma_{{\se}{\stau}} N_{\se}\;, \nonumber \\
\frac{dN_{\se}}{dt}&=&-\Gamma^{\se}_{\rm total}
N_{\se} \;,\nonumber \\
\frac{dN_{e}}{dt}&=&
\Gamma^{\se}_{FC} N_{\se} +
\Gamma^{\stau}_{FV,e} N_{\stau}\;, \nonumber
\\
\frac{dN_{\tau}}{dt}&=&
\Gamma^{\stau}_{FC} N_{\stau}+
\Gamma^{\se}_{FV,\tau} N_{\se}\;, \nonumber
\\\frac{dN_{\mu}}{dt}&=&
\Gamma^{\stau}_{FV,\mu} N_{\stau}+
\Gamma^{\se}_{FV,\mu} N_{\se} 
\label{diff-eqs}\;,
\eea 
where $N_{e,\mu,\tau}$ are the numbers of leptons coming directly from
gravitational two-body decays.

A fraction of taus will decay into muons and electrons with branching
ratios $BR(\tau^-\rightarrow \mu^- \;{\bar \nu_{\mu}} \;\nu_{\tau})=17.36
\pm 0.06\%$ and $BR(\tau^-\rightarrow e^- \;{\bar \nu_{e}}
\nu_{\tau})=17.84 \pm 0.06\%$~\cite{Eidelman:2004wy}.  These decay
products could mask the muons and electrons coming from the lepton
flavour violating decays ${\stau}\rightarrow \mu \; \psi_{3/2}$,
${\stau}\rightarrow e\; \psi_{3/2}$, and represent a source of
background. This background can be distinguished from the signal
by looking at the energy spectrum: the leptons from the flavour
conserving tau decay present a continuous energy spectrum, in stark
contrast with the leptons coming from the two body gravitational
decay, whose energies are sharply peaked at $E_0 =
(m^2_{{\stau},\,{\se}}+ m^2_{\mu,\,e}-m^2_{3/2})/(2
m_{{\stau},\,{\se}})$.  It is easy to check that only a very small
fraction of the electrons and muons from the tau decay have energies
close to this cut-off energy.  For instance, for the typical 
energy resolution of an electromagnetic calorimeter,
$\sigma\simeq 10\%/\sqrt{E({\rm GeV})}$, only
$2\times 10^{-5}$ of the taus with energy 
$E_0\sim 100$GeV will produce electrons
with energy $\simeq E_0$, within the energy 
resolution of the detector, which could be mistaken for electrons coming
from the LFV stau decay (see Appendix \ref{appendix-A} for
details). In the following, we will include this source of
background for completeness using the number $2\times 10^{-5}$, 
but in many instances it is negligible. For the muon
signal, however, this may be problematic since the measurement of the
muon energy is experimentally difficult. For simplicity, in the
following, we will concentrate on the search for $\stau\to e \;
\psi_{3/2}$.
                                                              
Some additional comments on possible experimental setups are in order.
To measure the temporal distributions of the events
precisely, the stopper itself had better be at the same time an active
detector~\cite{Hamaguchi:2004df}.  It measures the time when the NLSP
is stopped ($t_{\rm stop}$) and the time when it decays ($t_{\rm
decay}$) for each individual NLSP. By taking a distribution of the
($t_{\rm decay} - t_{\rm stop}$) for the samples, one can measure the
lifetime. In the following, we will assume that $\Gamma^{\stau}_{\rm
total} \simeq \Gamma(\stau\to \tau \; \psi_{3/2}) \ll
10^{-17}~\mathrm{GeV}$, or equivalently $\tau_{\stau}\gg
100~\mathrm{nsec}$, or $c\tau_{\stau}\gg 10~\mathrm{m}$. Otherwise
staus would decay inside the main detector and it would not be
possible to collect them in sufficiently large amounts.  This is the
case if $m_{3/2}\gg
1~\mathrm{keV}\times(m_{\stau}/100~\mathrm{GeV})^{5/2}$.
                                                                               
Concerning the production, there are roughly speaking two
possibilities. At the LHC, the numbers of produced $\stau$s,
$\se$s, (and $\smu$s) will be almost the same, since they
are mostly generated by cascade decays of squarks and gluinos. On the
other hand, at the Linear Collider, the numbers can be
different. Especially, at the $e^-e^-$ mode, suggested in
\cite{Hamaguchi:2004df} to produce many slow sleptons,
initially only $\se$s are produced.
                                                                              
Now let us discuss possible scenarios.
The total number of electrons detected can be obtained
solving eqs.(\ref{diff-eqs}). After a sufficiently long
observation time, and ignoring negligible backgrounds,
the solution can be simplified to
\bea
N_e\simeq N_{\se}\; BR({\se}\rightarrow e \;\psi_{3/2}) 
+\left(N_{\stau} +N_{\se} \;\frac{\Gamma_{\se \stau}}
{\Gamma^{\se}_{\rm total}}\right)
BR({\stau}\rightarrow e \;\psi_{3/2}).
\label{Nelec}
\eea
The first term is the most important source of background, which
becomes relevant when $N_{\se} \;BR({\se}\rightarrow e
\;\psi_{3/2})\gsim 1$.  This is going to arise only when the following
three conditions occur simultaneously:
\begin{itemize}
\item{$\se \rightarrow \stau \;e \;\tau$ is kinematically
closed, which requires $m_{\se}-m_{\stau}<m_{\tau}$.
}
\item{$\Gamma({\se}\rightarrow {\stau}\; e \; e)\lsim 
N_{\se}\;\Gamma({\se}\rightarrow e \;\psi_{3/2})$, which requires tiny 
lepton flavour violation. For instance, for the  
parameters of Fig.\ref{fig-gamma} and $\tan\beta=3$,
$A_0=0$, this condition requires 
$|U_{11}|^2+|U_{14}|^2  \lsim 10^{-14}$ for $N_{\se}=10^4$.
}
\item{$\Gamma({\se}\rightarrow {\stau}\; {\bar \nu_{\tau}}\; \nu_e)
\lsim N_{\se}\; \Gamma({\se}\rightarrow e \;\psi_{3/2})$. This condition
is the weakest among all. For the same parameters as above,
this condition requires $m_{3/2}\lsim 60$ GeV for $N_{\se}=10^4$.
}
\end{itemize}

\begin{figure}[t!]
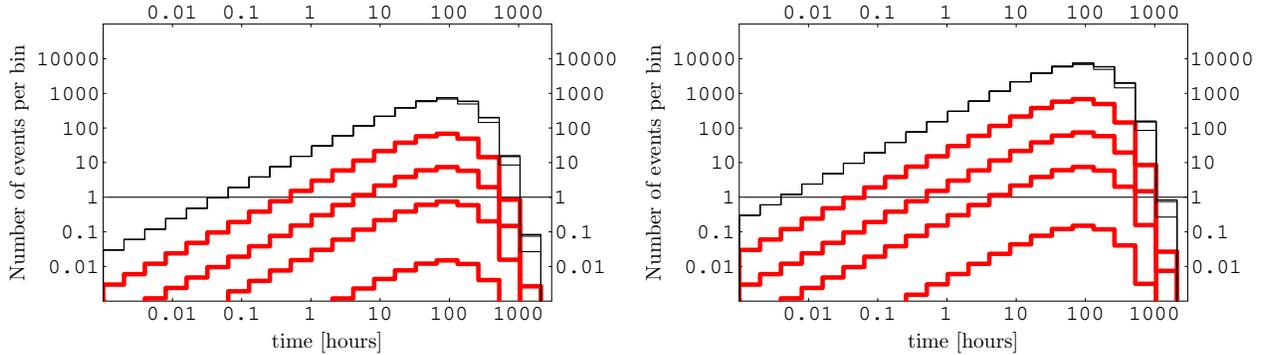

  \centerline{
    \scalebox{0.8}{\includegraphics{LHC3000-LFV.epsi}}
    ~
    \scalebox{0.8}{\includegraphics{LC30000-LFV.epsi}}
  }
  \caption{Number of electrons (thick red lines) and taus (thin black
  lines) per logarithmic time bin, $[t,2t]$, for the case in which only
  staus are present in the sample. In the left plot we show a typical case
  for the LHC $N_{\stau}=3000$, while in the right plot 
  for a $e^- e^-$ Linear Collider, $N_{\stau}=30000$.  
  We take as representative decay rates
  $\Gamma(\stau\to \tau\; \psi_{3/2})= \Gamma(\se\to
  e\; \psi_{3/2})=(100~\mathrm{hours})^{-1}$. The different
  curves correspond to different amounts of lepton flavour violation:
  $\Gamma(\stau \rightarrow e \; \psi_{3/2}) /\Gamma(\stau \rightarrow
  \tau \; \psi_{3/2})=0.1$, 0.01, 0.001 and 0 from
  the top to the bottom. The number of taus hardly differs when
  lepton flavour violation is switched on. Note that the background
  from the tau decay (the lowest line) is negligible.}
  \label{fig-LFV}
\end{figure}

When any of these three conditions fails, the backgrounds
are essentially negligible, therefore any electron 
observed in the detector would be a clear indication 
for LFV. 
To estimate the prospects for detection of flavour violation, we will
consider the cases of the LHC and the $e^- e^-$ Linear Collider (LC),
and we will assume that particle identification is very good in the
stopper-detector. The temporal distribution of electrons 
is shown in Fig.\ref{fig-LFV} for
different amounts of lepton flavour violation. 
As typical numbers for the LHC, we take
$N_{\stau}({\rm init.})=N_{\smu}({\rm init.})=
N_{\se}({\rm init.})=1000$, so that after the selectrons
and smuons have decayed, the sample consists of 3000 staus.
The total number of electrons 
observed over the whole observation time is 
273, 29.7, 3.0 and 0.06 for 
$\Gamma(\stau \rightarrow e \; \psi_{3/2}) /\Gamma(\stau \rightarrow
\tau \; \psi_{3/2})=0.1$, 0.01, 0.001 and 0,
respectively. On the other hand, for the LC we take as 
a representative possibility $N_{\stau}({\rm init.})=0,
\,N_{\smu}({\rm init.})=0,\, N_{\se}({\rm init.})=30000$, 
which lead to $N_{\stau}=30000$ after selectrons decay.
In this case, the number of events are 2727, 297, 30.0 and 0.6 
(notice that the number of events from the
background is always smaller than one).

Inversely, if no electron is observed, stringent bounds on lepton
flavour violation would follow. To be precise, for the LHC
we estimate that if no electrons are observed, the bound
\bea
|U_{11}|^2+|U_{14}|^2 &\lsim& 8\times10^{-4}
\quad {\rm for}\, N_{\stau}=3000,
\eea
would follow at 90\% confidence level.
Analogously, for the LC, if no electrons are observed,
the corresponding bound would be
\bea
 |U_{11}|^2+|U_{14}|^2\lsim 8\times 10^{-5}
\quad {\rm for}\, N_{\stau}=30000,
\eea
at 90\% confidence level. Hence, mixing angles in 
the slepton sector as small as $\sim 3\times 10^{-2}~(9\times 10^{-3})$ 
could be probed at the 90\% confidence level
if $3\times 10^3$ ($3\times 10^4$) sleptons could be collected.
This is the main result of this paper.

It is interesting to mention that when left-right mixing is not
too large (in particular, for small or moderate $\tan\beta$),
the NLSP consists mainly of a right-handed stau. Therefore,
in their decays we would be probing flavour violation in the
right-handed sector, which is a rather elusive sector. 
Lepton flavour violating processes generated radiatively, such as 
$\tau\rightarrow e\; \gamma$, are very insensitive
to the right-handed sector, resulting in very poor constraints
for the right-handed mixing angles \cite{Masina:2002mv}.
In consequence, the tree-level 
flavour violating stau decays studied in this paper 
could offer a unique opportunity to probe this sector,
if this scenario is realized in nature.

Let us discuss now the results in the cases where
the three conditions above are simultaneously fulfilled.
If there is no LFV, or the amount of LFV is so small
that $\Gamma({\se}\rightarrow {\stau}\; e \; e)\ll
\Gamma({\se}\rightarrow e \;\psi_{3/2})$, then selectrons
decay mainly via ${\se}\rightarrow {\stau}\; {\bar \nu_{\tau}}\; \nu_e$ 
or ${\se}\rightarrow e \;\psi_{3/2}$. Since the decay rates
of those modes are very much suppressed, selectrons reach
the stopper-detector before decaying, and two different
scenarios could arise depending on the relative
size of ${\se}\rightarrow {\stau}\; {\bar \nu_{\tau}}\; \nu_e$ 
and ${\se}\rightarrow e \;\psi_{3/2}$.

\begin{figure}[t!]
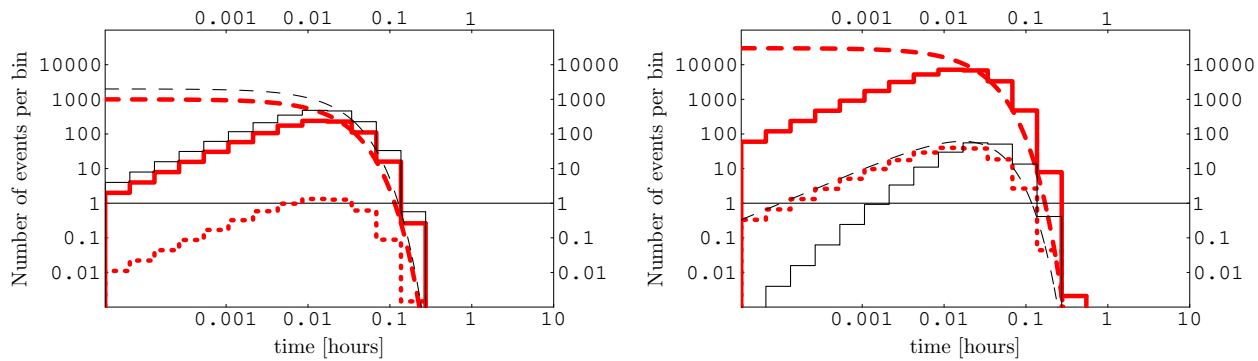

  \centerline{
    \scalebox{0.8}{\includegraphics{LHC20001000-1m-3h.epsi}}
    ~
    \scalebox{0.8}{\includegraphics{LC30000-1m-3h.epsi}}
  }
  \caption{Number of events per logarithmic time bin, $[t,2t]$, for
    a typical case in which the sample consists of both selectrons
    and staus, and $\Gamma(\se\to \stau\;\nu_e\;\bar{\nu}_{\tau}) \ll
    \Gamma(\se\to e \; \psi_{3/2})$. In the left plot we show the
    number of events for $N_{\stau}({\rm init.})=2000$ and 
    $N_{\se}({\rm init.})=1000$,
    while in the right plot $N_{\stau}({\rm init.})=0,\,N_{\se}({\rm
    init.})=30000$. The different decay rates are $\Gamma(\stau\to
    \tau\; \psi_{3/2})= \Gamma(\se\to
    e\; \psi_{3/2})=(1~\mathrm{min})^{-1}$ and $\Gamma(\se\to
    \stau\;\nu_e\;\bar{\nu}_{\tau})=(3~\mathrm{hours})^{-1}$.  The thick
    red and the thin black histograms represent the number of electrons
    and taus, respectively, whereas the dotted red histogram represent
    the (invisible) events $\se\to\stau$. We also show the number of
    stau (thick red dashed line) and selectron (thin black dashed
    line).}
  \label{fig-LFC-3}
\end{figure}

If  $\Gamma(\se\to \stau\;\nu_e\;\bar{\nu}_{\tau}) \ll
\Gamma(\se\to e \; \psi_{3/2})$, the conversion 
of selectrons into staus is very slow, and the dominant 
decay channel of the selectrons is the
gravitational one, with the same lifetime as the staus. 
Therefore, in the detector we would observe electrons and taus,
presenting a temporal distribution with peaks at very
similar times. A clear signature for this scenario
would be the detection of a large amount of electrons
in the detector, as many as selectrons were produced
initially, so this case is clearly distinguishable
from the other ones. In Fig.\ref{fig-LFC-3} we show the electron
and tau distribution for typical parameters corresponding
to this scenario. 
If this possibility is realized in nature, it would be possible to
measure the stau and the selectron lifetimes from the events
$\stau\to \tau \; \psi_{3/2}$ and $\se\to e \; \psi_{3/2}$, and accordingly 
test the universality of the gravitino coupling.

On the other hand, it could happen that
$\Gamma({\se}\rightarrow {\stau}\; {\bar \nu_{\tau}}\; \nu_e)\gg
\Gamma(\se\to e \; \psi_{3/2})$ (while preserving the condition 
$\Gamma({\se}\rightarrow {\stau}\; {\bar \nu_{\tau}}\; \nu_e)
\lsim N_{\se}\;\Gamma({\se}\rightarrow e \;\psi_{3/2}))$.
In this case, one could see both electron events and tau events in the
detector, with the peaks of their temporal distribution occurring
at different times, corresponding to the two different lifetimes
$\tau_{\stau}$ and $\tau_{\se}$.
This case is potentially dangerous for the study of lepton
flavour violation. In contrast to the case 
with $\Gamma(\se\to \stau\;\nu_e\;\bar{\nu}_{\tau}) \ll
\Gamma(\se\to e \; \psi_{3/2})$, now an important fraction of selectrons
could decay into staus, so the number of electron events in the
detector could be small. Therefore, just from counting
the total number of events, this scenario could be 
mistaken with the scenario with small flavour violation,
that would also yield a small number of electrons in the detector.
Nevertheless, it could be possible to discriminate this scenario
from the one with small flavour violation
by looking at the temporal distribution of events. This is
shown in Fig.\ref{fig-LFC-2} for typical parameters corresponding to
this scenario. Comparing with Fig.\ref{fig-LFV}, it is apparent that
in this case the peak of the electron distribution appears at
earlier times, and clearly displaced from the peak of the tau 
distribution. Another possibility to discriminate between 
these two scenarios would be searching the two electrons emitted
in the flavour violating selectron decay mediated by neutralinos,
$\se \rightarrow \stau \;e \;e$. Notice that these decays could
offer an alternative method for detecting flavour violation,
although the analysis of the backgrounds is much more 
involved. A further study along this lines would 
be certainly interesting, but lies
beyond the scope of this paper. Detailed studies of lepton 
flavour violation via $\se \rightarrow \stau \;e \;e$ will
be presented elsewhere.

\begin{figure}[t!]
  \centerline{
    \scalebox{0.8}{\includegraphics{LHC20001000-100h-3h.epsi}}
    ~
    \scalebox{0.8}{\includegraphics{LC30000-100h-3h.epsi}}
  }
  \caption{The same as Fig.\ref{fig-LFC-3} but for a case
    with $\Gamma(\se\to \stau\;\nu_e\;\bar{\nu}_{\tau}) \gg
    \Gamma(\se\to e \; \psi_{3/2})$. We take as representative decay
    rates  $\Gamma(\stau\to \tau\; \psi_{3/2})= \Gamma(\se\to
    e\; \psi_{3/2})=(100~\mathrm{hours})^{-1}$ and $\Gamma(\se\to
    \stau\;\nu_e\;\bar{\nu}_{\tau})=(3~\mathrm{hours})^{-1}$.}
  \label{fig-LFC-2}
\end{figure}

\section{Conclusions}

In this paper we have discussed the prospects for
observing lepton flavour violation 
in the decay of the lightest charged slepton,
in scenarios where the gravitino is the lightest
superparticle. In this class of scenarios,
renormalization group analyses of the
masses of the superparticles and different cosmological arguments,
point to the possibility that 
the lightest charged slepton, generically a right-handed stau,
is the next-to-lightest superparticle. Therefore, it
can only decay gravitationally into charged leptons
and gravitinos, with very long lifetimes due to the
gravitational suppression of the interaction. This would allow
the collection of staus and the clean observation of their
decay products. The normal (flavour conserving) decay 
mode of staus would yield taus, in consequence, the observation of
electrons and muons in the detector would be an indication
for lepton flavour violation. 

We have studied the different sources of
backgrounds and identified the potentially most disturbing 
background, namely the possibility that selectrons are as long
lived as staus, so that the electrons from their flavour conserving
decays could be mistaken for electrons from the flavour violating
stau decay. Nevertheless, we have shown that if flavour violation is
large enough to be observed in these experiments, the selectron
decay channel $\se\to\stau \;e\; e$ is very efficient. Therefore,
selectrons are never long lived enough to represent an important
source of background. We have remarked the interesting double role
that flavour violation plays in this experiment, both as object
of investigation and as crucial ingredient for the success
of the experiment itself. 

We have estimated that in the LHC or the future Linear
Collider it would be possible
to probe mixing angles in the slepton sector
down to the level of 
$\sim 3\times 10^{-2}~(9\times 10^{-3})$ at 90\% confidence level if
$3\times 10^3~(3\times 10^4)$  staus could be collected.
It is important to stress that this experiment 
probes directly flavour violation in the right-handed sector,
which is a sector rather difficult to constrain
from radiative rare decays, such as $\tau\rightarrow e\; \gamma$.

\section*{Acknowledgements}
We would like to thank the organizers of SUSY04, Tsukuba, Japan, June
17-23, 2004, where this work has been initiated. We are grateful to
Concha Gonz\'alez-Garc{\'\i}a and Mihoko M.~Nojiri for helpful discussions.

\appendix
                                                                               
\section{Electron spectrum from tau decay}
\label{appendix-A}

We would like to review here the electron
spectrum coming from the decay of 
taus in flight \cite{Gomez-Cadenas:1988sm}
(the muon spectrum is completely analogous). 
We assume that the NLSPs are stopped in the laboratory
frame, so the taus from NLSP decay have an energy
$E_{\tau} =(m^2_{\stau}+m^2_{\tau}-m^2_{3/2})/(2 m_{\stau})$.  
Generically, and as long as $m_{\stau}-m_{3/2}\gg m_{\tau}$, 
the outgoing leptons from NLSP decay are 
ultrarelativistic. Therefore, the cut-off of the 
energy spectrum of the electrons from tau decay, 
$E_{\tau}$, is very similar to the energy of the electrons from
the LFV stau decay, $E_0$, defined after eq.(\ref{diff-eqs}).

Taus decay into electrons with a continuous
spectrum, whose peak lies at
\bea
E_{peak}=\frac{1}{2} (E_{\tau}-\sqrt{E^2_{\tau}-m^2_{\tau}})
\simeq \frac{m_{\tau}^2}{4 E_{\tau}},
\eea
where we have used $E_{\tau}\gg m_{\tau}$ and we have neglected
the mass of the electron.
It is important to note that as the taus become ultrarelativistic,
the peak of the electron spectrum is shifted to low energies. Therefore,
the number of electrons with energies close to $E_0$
coming from the tau decay is tiny, which reduces enormously
the background for the lepton flavour violating decay.
For energies smaller than $E_{peak}$, the electron spectrum
is described by
\bea
\frac{d\Gamma}{dE_{e}}=\frac{G_F^2}{18 \pi^3} 
\frac{\sqrt{E_{e}^2-m_{e}^2}}{E_{\tau}}
\left[
2E_{e}^2(m^2_{\tau}-4 E_{\tau}^2)
+\frac{9}{2} E_{e} E_{\tau} (m_{e}^2+m_{\tau}^2)
+m_{e}^2(2 E_{\tau}^2-5 m_{\tau}^2)
\right]
\eea
while for energies larger than $E_{peak}$, the 
corresponding expression reads
\bea
\frac{d\Gamma}{dE_{e}}=\frac{G_F^2}{192 \pi^3} 
\frac{\Omega}{E_{\tau}\sqrt{E_{\tau}^2-m_{\tau}^2}}
\left[-\frac{4}{3} \Omega^2+ \Omega(m_{e}^2+m_\tau^2)
+2(m_{e}^2-m_{\tau}^2)^2 \right],
\eea
where $\Omega=m_{\tau}^2+m_{e}^2-2E_{\tau} E_{e}+
2\sqrt{E_{\tau}^2-m_{\tau}^2} \sqrt{E^2_{e}-m_{e}^2}$ and $E_{e}$
is the energy of the outgoing electron in the laboratory frame.

In the region of interest, the region close to the cut-off $E_0$, 
this formula is simplified to:
\bea
\frac{d\Gamma}{d E_{e}} \simeq \frac{G_F^2}{192 \pi^3} 
\frac{m^6_{\tau}}{E^2_{\tau}} \left[\frac{5}{3}
-3 \left(\frac{E_{e}}{E_{\tau}} \right)^2
+\frac{4}{3} \left(\frac{E_{e}}{E_{\tau}}\right)^3 \right],
\label{spectrum}
\eea
where we have neglected the mass of the outgoing electron, 
and we have taken the limit $E_{\tau} \gg m_{\tau}$.
It is convenient to express the energy of the electrons
relative to the cut-off energy, defining $x=E_{e}/E_{\tau}$.
With this definition, $0<x\lsim 1-m_{\tau}^2/4E_{\tau}^2$. 
Therefore, the energy spectrum in the region of interest 
can be rewritten as:
\bea
\frac{d \Gamma}{dx} \simeq \frac{G_F^2}{192 \pi^3} 
\frac{m^6_{\tau}}{E_{\tau}} f(x),
\eea
where $f(x) \equiv \frac{5}{3} -3 x^2 +\frac{4}{3} x^3$
is the fraction of electrons with energy $x$.

We are interested in electrons with energies close to the
cut-off $E_0$, {\it i.e.} with energies in the range
$1-\Delta x<x\lsim 1-m_{\tau}^2/4E_{\tau}^2$, 
being $\Delta x \ll 1$.
The number of electrons in this energy bin is:
\bea
N^{LFC}_{e}=
N_{\stau} BR(\stau\rightarrow \tau \; \psi_{3/2})
BR(\tau\rightarrow e\; {\bar \nu_{e}}\; \nu_{\tau})   
\int_{1-\Delta x}^{1-\frac{m^2_{\tau}}{4E_{\tau}^2}} dy
\int_0^{1-\frac{m^2_{\tau}}{4E_{\tau}^2}} dz f(z)
\frac{e^{-\frac{(z-y)^2}{2\sigma^2}}}{\sigma \sqrt{2 \pi}}.
\eea
For a typical electromagnetic calorimeter, the energy
resolution is $\sigma \simeq 10\%/\sqrt{E({\rm GeV})}$.
Therefore, the fraction of electrons in that energy bin, that
can be mistaken for electrons coming from the LFV stau decay is
$N^{LFC}_{e}\simeq 2\times 10^{-5} N_{\stau}$
for $E_\tau \simeq 100~\mathrm{GeV}$.
For the number of NLSPs that we are using to illustrate
our results, the number of electrons from this source
of background turns out to be negligible in most instances.

\section{Probing $R$-parity violation in slepton NLSP scenarios}
                                                                               
In the body of the paper we have assumed that the $R$-parity is
conserved. In this appendix we briefly discuss possible probe of
R-parity violations via long-lived charged sleptons. Clearly, the long
lifetime of the NLSP slepton leads to a severe bound on the $R$-parity
violating couplings. Let us concentrate on the following trilinear
$R$-parity violating operators in the superpotential
\begin{eqnarray}
  W_{RpV} &=&
  \frac{1}{2}\lambda^{ijk}L_i L_j E^c_k
  +
  \lambda'^{ijk}L_i Q_j D^c_k,
\end{eqnarray}
and assume that the NLSP is mainly the lighter stau, $\stau_1 =
\cos\theta_{\stau}\stau_R + \sin\theta_{\stau}\stau_L$. If we observe
that the stau decays mainly into tau and gravitino (missing particle)
with lifetime $\tau_{\stau}$, one obtains the following bounds
\begin{eqnarray}
  \left|\lambda^{3jk}\sin\theta_{\stau}\right|^2,\,
  2\left|\lambda^{ij3}\cos\theta_{\stau}\right|^2,\,
  3\left|\lambda'^{3jk}\sin\theta_{\stau}\right|^2
  \lsim
  10^{-30}\times \left(\frac{100~\mathrm{GeV}}{m_{\stau}}\right)
  \left(\frac{100~\mathrm{hours}}{\tau_{\stau}}\right),
\end{eqnarray}
where we have assumed that there is no cancellation between
couplings. Note that the bounds range by orders of magnitude depending
on the lifetime.
     
If on the contrary the $R$-parity violation would dominate the decay,
the signals would be (i) for $\lambda'^{3jk}$ coupling; two jets with
identical energies $E_{jet}=E_0=m_{\stau}/2$, (ii) for
$\lambda^{123}$; one lepton ($e$ or $\mu$) with energy $E_0$ plus
missing particle, with branching ratios being $BR(\stau\to
e\;\nu_\mu)=BR(\stau\to \mu\;\nu_e)$, (iii) for $\lambda^{3jk}$ ($k\ne
3$); one lepton ($l_k$, $k\ne 3$) with energy $E_0$ plus missing
particle, and (iv) for $\lambda^{3j3}$; one lepton ($\tau$ or $l_j$,
$j\ne 3$) with energy $E_0$ plus missing particle with $BR(\stau\to
\tau\;\nu_j)\sim BR(\stau\to l_j\;\nu_\tau)$.
                                               

\end{document}